\documentclass[pra,twocolumn, notitlepage,superscriptaddress,longbibliography]{revtex4-1}
\usepackage[english]{babel}
\usepackage{amsmath,amssymb,bbm,mathrsfs,bm,mathtools,braket,color,graphicx,comment,textcomp,amsfonts,dsfont}
\usepackage[colorlinks,citecolor=blue,urlcolor=blue]{hyperref}

\AtBeginDocument{%
    \newwrite\bibnotes
    \def\bibnotesext{Notes.bib}
    \immediate\openout\bibnotes=\jobname\bibnotesext
    \immediate\write\bibnotes{@CONTROL{REVTEX41Control}}
    \immediate\write\bibnotes{@CONTROL{%
    apsrev41Control,author="08",editor="1",pages="1",title="0",year="1"}}
     \if@filesw
     \immediate\write\@auxout{\string\citation{apsrev41Control}}%
    \fi
}%

\allowdisplaybreaks

\newcommand{\id}{\mathbbm{1}}
\newcommand{\Hc}{\mathrm{H.c.}}
\newcommand{\N}{\mathbbm{N}}

\newcommand{\peq}{\mathrel{\hphantom{=}}}

\newcommand{\abs}[1]{\left\lvert #1 \right\rvert}

\newcommand{\tr}{\mathop{\mathrm{tr}}}
\newcommand{\sinc}{\mathop{\mathrm{sinc}}}

\renewcommand{\Im}{\mathop{\mathrm{Im}}}

\newcommand{\kket}[1]{\left\lvert #1 \right\rangle_{\sharp}}
\newcommand{\bbra}[1]{\prescript{}{\sharp}{\left\langle #1 \right\rvert}}
\newcommand{\bbraket}[1]{\braket{ #1 }_{\sharp}}

\begin{document}

\title{Statistical periodicity in driven quantum systems: General formalism
  and application to noisy Floquet topological chains}

\author{Lukas M. Sieberer}

\email{lukas.sieberer@uibk.ac.at}

\affiliation{Department of Physics, University of California, Berkeley,
  California 94720, USA}

\affiliation{Center for Quantum Physics, University of Innsbruck, 6020
  Innsbruck, Austria}

\affiliation{Institute for Quantum Optics and Quantum Information of the
  Austrian Academy of Sciences, A-6020 Innsbruck, Austria}

\author{Maria-Theresa Rieder}

\affiliation{Department of Condensed Matter Physics, Weizmann Institute of Science, Rehovot 7610001, Israel}

\author{Mark H. Fischer}

\affiliation{Institute for Theoretical Physics, ETH Zurich, 8093 Zurich, Switzerland}

\author{Ion C. Fulga}

\affiliation{IFW Dresden, Helmholtzstr.~20, 01069 Dresden, Germany}

\begin{abstract}  
  Much recent experimental effort has focused on the realization of exotic
  quantum states and dynamics predicted to occur in periodically driven systems.
  But how robust are the sought-after features, such as Floquet topological
  surface states, against unavoidable imperfections in the periodic driving? In
  this work, we address this question in a broader context and study the
  dynamics of quantum systems subject to noise with periodically recurring
  statistics. We show that the stroboscopic time evolution of such systems is
  described by a noise-averaged Floquet superoperator. The eigenvectors and
  -values of this superoperator generalize the familiar concepts of Floquet
  states and quasienergies and allow us to describe decoherence due to noise
  efficiently.  Applying the general formalism to the example of a noisy Floquet
  topological chain, we re-derive and corroborate our recent findings on the
  noise-induced decay of topologically protected end states. These results
  follow directly from an expansion of the end state in eigenvectors of the
  Floquet superoperator.
\end{abstract}

\maketitle

\section{Introduction}
\label{sec:introduction}

The periodic modulation of a quantum system is a powerful tool to engineer
exotic, effectively static models. Even more intriguingly, it provides a pathway
to realize novel phases of matter without time-independent counterparts. Central
to these ideas is the existence of Floquet states that generalize the notion of
eigenstates of a static Hamiltonian to the periodically driven
setting~\cite{Eckardt2017}. In the basis of Floquet states, the stroboscopic
time evolution of a driven system resembles the undriven Hamiltonian dynamics,
albeit described by an effective Hamiltonian~\cite{Rahav2003, Rahav2003a,
  Goldman2014}. In particular, a system that is initialized in a Floquet
eigenstate, remains in this state.  \textit{Floquet engineering}, thus, amounts
to designing an effective Hamiltonian such that its spectrum and eigenstates,
i.e., the quasienergies and Floquet states, have the desired, e.g.,
topological~\cite{Kitagawa2010, Rudner2013}, properties.

A common feature of all such proposals is that a perfect periodicity of the
driving is required for a well-defined effective Hamiltonian and Floquet
eigenstates. This raises questions about the effects of imperfections in the
driving protocol. In a recent paper~\cite{Rieder2017}, we addressed this
question for the specific example of a Floquet topological chain with timing
noise. For perfectly periodic driving, this system hosts topologically protected
end states. Timing noise induces transitions between Floquet states and causes a
particle that is initialized in such an end state to decay into the
bulk. Interestingly, this decay is slowed down substantially if the bulk states
are localized. The noise in this case can be treated as a perturbation to the
perfectly periodic driving leading to dynamics captured by a discrete-time
Floquet-Lindblad equation (FLE).

Here, we broaden the scope of this investigation. We show that the stroboscopic
evolution of a driven system with stochastic periodicity, i.e., with random
fluctuations that obey periodically recurring statistics, is described by a
Floquet superoperator which directly generalizes the notion of a conventional
Floquet operator. This description is not limited to weak noise and does not even rely on the existence
of a perfectly periodic and noiseless limit. In this sense, it also applies to
systems which are driven exclusively by noise. Within this formalism, we 
re-derive and corroborate the results for a noisy Floquet topological chain as
reported in Ref.~\cite{Rieder2017}. Here, we obtain the decay of the end
state by calculating the eigenoperators and eigenvalues of the noise-averaged
Floquet superoperator.

The remainder of the paper is organized as follows: We begin in
Sec.~\ref{sec:stat-peri} by defining the notion of statistical periodicity that
is underlying our work and we introduce key concepts such as the Floquet
superoperator. The formalism to describe systems that are periodic on average is
developed in Sec.~\ref{sec:noisy-floq-syst} for the example of a Floquet system
that is perturbed by timing noise. We show in
Sec.~\ref{sec:fully-random-driving} how this approach captures fully random
driving with temporally periodic statistics. Section~\ref{sec:appl-noisy-floq}
illustrates the theory by applying it to the noisy Floquet topological chain we
studied previously in Ref.~\cite{Rieder2017}. Finally, we discuss our results
and future directions in Sec.~\ref{sec:conclusions-outlook}.

\section{Statistical periodicity}
\label{sec:stat-peri}

In this paper, we study the dynamics of quantum systems subject to stochastic
driving protocols with periodically recurring statistics. More specifically, we
focus on driving protocols which can be defined in terms of elementary driving
cycles that are applied repeatedly, but with random and statistically
independent parameter fluctuations from cycle to cycle. Since this definition is
rather general, we find it worthwhile to briefly mention the perhaps simplest
incarnation of such a driving protocol. In this simple example, an elementary
cycle consists of applying a (constant) Hamiltonian $H_1$ for a duration of
$T_1$ followed by the application of (another) Hamiltonian $H_2$ for a duration
of $T_2$, where $T_1$ and $T_2$ are random and are drawn from the same
distribution in each individual cycle. One of the central goals of this paper is
to derive an evolution equation describing the noise-averaged dynamics of such
systems.

To this end, we first consider the evolution under a specific realization of the
stochastic drive, and we denote the quantum state of the system after $n$
elementary driving cycles by $\ket{\psi_n}$. The system's evolution during one
driving cycle is given by
\begin{equation}
  \label{eq:psi_evolution}
  \ket{\psi_{n + 1}} = U_{{\rm F},n + 1} \ket{\psi_n},
\end{equation}
where the subscript F suggests that for deterministic driving, i.e., when
$U_{ {\rm F}, n} = U_{\rm F}$ for all $n$, we recover a standard Floquet problem (if $U_{\rm F}$ is
generated by a Hamiltonian with non-trivial time dependence). By taking the
average over different realizations of the stochastic drive, which we indicate
by an overbar hereafter, we obtain the density matrix
$\rho_n = \overline{\ket{\psi_n} \bra{\psi_n}}$. Working with the density matrix
instead of the noise-averaged pure state allows us to directly calculate
physical observables as
\begin{equation}
  \label{eq:observable-average}
  \overline{\langle O_n \rangle} = \overline{\braket{\psi_n | O | \psi_n}} = \tr
  \! \left( O \overline{\ket{\psi_n} \bra{\psi_n}} \right) = \tr(O \rho_n).
\end{equation}
The evolution equation for $\rho_n$ can be cast as a generalization of the
familiar relation $\ket{\psi_{n + 1}} = U_{\rm F} \ket{\psi_n}$ for noiseless
Floquet systems, where $U_{\rm F}$ is the usual Floquet operator describing the
time evolution over one period. To derive this evolution equation, we start by
expressing $\rho_{n + 1}$ as the average of Eq.~\eqref{eq:psi_evolution},
\begin{equation}
  \rho_{n + 1} = \overline{\ket{\psi_{n + 1}} \bra{\psi_{n + 1}}} =
  \overline{U_{ {\rm F}, n + 1} \ket{\psi_n} \bra{\psi_n} U_{ {\rm F}, n +
      1}^{\dagger}}. 
\end{equation}
Under the assumption of statistical independence of fluctuations which occur in
different driving cycles, the average in the above expression factorizes and can
be taken separately over $U_{\mathrm{F}, n + 1}$ and $\ket{\psi_n}$. We thus
obtain
\begin{equation}
  \label{eq:caligraphic-F}
  \rho_{n + 1} = \overline{U_{ {\rm F}, n + 1} \rho_n U_{ {\rm F}, n + 1}^{\dagger}} = \mathcal{F} \rho_n.
\end{equation}
This relation defines the noise-averaged Floquet superoperator $\mathcal{F}$. We
note that while we introduce $\mathcal{F}$ to describe the dynamics of systems
subject to stochastic driving, the concept of a Floquet superoperator also
arises in periodically driven systems in the presence of (Markovian)
dissipation~\cite{Haake2010}. To illustrate the basic properties of the Floquet
superoperator $\mathcal{F}$ it is useful to first consider the case of perfectly
periodic driving described by a Floquet operator $U_{\rm F}$ such that
$\mathcal{F} = \mathcal{U}_{\rm F}$ and
\begin{equation}
  \label{eq:caligraphic-U-F}
  \rho_{n + 1} = \mathcal{U}_{\rm F} \rho_n = U_{\rm F} \rho_n U_{\rm F}^{\dagger}.
\end{equation}
Then, the eigenvalues and eigenoperators of $\mathcal{U}_{\rm F}$ can be obtained
immediately from an eigenbasis of the Floquet operator $U_{\rm F}$. For the states
$\ket{\alpha}$ which satisfy
$U_{\rm F} \ket{\alpha} = e^{-i T \varepsilon_{\alpha}} \ket{\alpha}$, where $T$ is
the period of the drive and $\varepsilon_{\alpha}$ the quasienergy of the state
$\ket{\alpha}$, we find
\begin{equation}
    \label{eq:UF}
    \mathcal{U}_{\rm F}(\ket{\alpha} \bra{\beta}) = U_{\rm F} \ket{\alpha} \bra{\beta}
    U_{\rm F}^{\dagger} = e^{-i T \left( \varepsilon_{\alpha} - \varepsilon_{\beta}
    \right)} \ket{\alpha} \bra{\beta}, 
\end{equation}
i.e., the eigenoperators of $\mathcal{U}_{\rm F}$ take the form
$\ket{\alpha} \bra{\beta}$ and the corresponding eigenvalues are determined by
the differences of quasienergies $\varepsilon_{\alpha} - \varepsilon_{\beta}$.
Unitarity of the perfectly periodic dynamics guarantees that the eigenvalues of
$\mathcal{U}_{\rm F}$ have unit modulus. This does not apply if the dynamics is
generated by a stochastic driving protocol. For instance, timing noise as
considered in Ref.~\cite{Rieder2017} and Sec.~\ref{sec:timing-noise} below
induces transitions between the Floquet states $\ket{\alpha}$, causing the
system to eventually heat up to infinite temperature, i.e., for $n \to \infty$
the state becomes fully mixed, $\rho_n \to \id/D$, where $D$ is the dimension of
the Hilbert space. In the spectrum of $\mathcal{F}$, this is reflected in $\id$
being the unique eigenoperator corresponding to the eigenvalue $1$, i.e.,
$\mathcal{F} \id = \id$. Writing the eigenvalues of $\mathcal{F}$ as
$e^{-i \lambda}$, noise-induced decoherence implies that all ``eigenphases''
$\lambda \neq 0$ corresponding to non-stationary states acquire finite imaginary
parts $\Im(\lambda) < 0$. Thus, the spectrum of $\mathcal{F}$ reveals that at
long times noisy Floquet systems generically heat up to a featureless
infinite-temperature state, but remnants of the properties of the system for
perfectly periodic driving can survive in the dynamics. The latter can also be
described efficiently by expanding the system's state in eigenoperators of
$\mathcal{F}$. We pursue this strategy for the example of a noisy Floquet
topological chain in Sec.~\ref{sec:appl-noisy-floq}. First, however, we derive
formal expressions for the Floquet superoperator for noisy Floquet systems and
for systems which are subject to purely random driving in
Secs.~\ref{sec:noisy-floq-syst} and~\ref{sec:fully-random-driving},
respectively.

\section{Noisy Floquet systems}
\label{sec:noisy-floq-syst}

To connect to the familiar physics and concepts of periodically driven (Floquet)
systems, we first introduce stochastic driving as a perturbation. That is, we
consider a slight imperfection causing random fluctuations around a periodic
driving protocol. Motivated by much recent work on Floquet systems, we consider
piecewise constant driving, i.e., the time-dependence of the Hamiltonian is a
succession of sudden quenches interluded by phases during which the Hamiltonian
is kept constant. Moreover, to not burden the discussion with unnecessary
complications, we develop the formalism for the simplest case of binary driving
that is defined in terms of just two Hamiltonians $H_{1, 2}$ which are
alternated. The generalization of our considerations to a driving protocol that
comprises more than two steps is straightforward and summarized in
Sec.~\ref{sec:multi-phase-pc}.

\subsection{Timing noise}
\label{sec:timing-noise}

Our starting point is a perfectly periodic driving protocol described by the
Hamiltonian
\begin{equation}
  \label{eq:binary-driving}
  H(t) =
  \begin{cases}
    H_1, & n T \leq t < n T + T_1, \\
    H_2, & n T + T_1 \leq t < \left( n + 1 \right) T,
  \end{cases}
\end{equation}
i.e., $H(t) = H_1$ is kept constant for a time span of length $T_1$, and then
switched to $H_2$ which is applied for $T_2$. The cycle of duration
$T = T_1 + T_2$ is then repeated periodically, and the integer $n$ counts the
number of elapsed driving cycles. The evolution of the system during one driving
period is described by the Floquet operator $U_{\rm F}$, which is given by
\begin{equation}
  \label{eq:UF_binary}
  U_{\rm F} = \mathsf{T} e^{-i \int_0^T dt \, H(t)} = U_2 U_1 = e^{-i T_2 H_2} e^{-i
    T_1 H_1},
\end{equation}
where $\mathsf{T}$ denotes time ordering. A special case of this driving
protocol arises in the limit when $H_2$ is applied as a short pulse of strength
$\lambda$ and duration $T_2 \to 0$. This limit defines the class of periodically
``kicked'' systems with time-dependent Hamiltonian
\begin{equation}
  \label{eq:H_kicked}
  H(t) = H_1 + \lambda \sum_{n \in \N} \delta(t - n T) H_2,
\end{equation}
giving rise to a Floquet operator of the form
$U_{\rm F} = e^{-i \lambda H_2} e^{-i T H_1}$. The Floquet operator determines the
stroboscopic time evolution of the state of the system: At multiples of the
driving period $T$, the system's state is given by
$\ket{\psi_n} = \ket{\psi(n T)} = U_{\rm F}^n \ket{\psi_0}$ for the initial state
$\ket{\psi_0}$. By diagonalizing
$U_{\rm F} \ket{\alpha} = e^{-i T \varepsilon_\alpha} \ket{\alpha}$, we can thus describe
the system in terms of its Floquet eigenstates $\ket{\alpha}$ and their
associated quasienergies $\varepsilon_\alpha$.

The ideal scenario of perfectly periodic driving outlined thus far---and
extensions to more elaborate driving schemes---form the basis of a great number
of proposals to design Floquet quantum matter. In any realistic experimental
implementation, however, noise and imperfections cannot be eliminated
completely. A case in point are experiments that demonstrate Floquet topological
insulators in photonic waveguides~\cite{Mukherjee2017, Maczewsky2017}, where
time evolution of a quantum system is emulated by the propagation of light in
the waveguide~\cite{Longhi2009, Szameit2010}. Consequently, fabrication defects
in the waveguide amount to timing noise in the emulated quantum dynamics. A
simple model for this type of noise replaces the duration $T_i$ for which the
Hamiltonian $H_i$ in Eq.~\eqref{eq:binary-driving} is applied during the $n$th
driving cycle by $T_{in} = T_i + \tau_{in}$. Noise in $T_{in}$ is
incorporated in the random number $\tau_{in}$, which has vanishing mean,
$\overline{\tau_{in}} = 0$ and fluctuations given by
$\overline{\tau_{in}^2} = \tau^2$. Further, we assume that the time shifts in
different parts of a single cycle as well as in different cycles are
uncorrelated, i.e.,
$\overline{\tau_{in} \tau_{jn'}} = \tau^2 \delta_{ij} \delta_{nn'}$. This
assumption is reasonable in experiments in which the binary driving is realized
by sudden quenches of the Hamiltonian parameters that suffer from imperfections.
The time-dependent Hamiltonian is then given by
\begin{equation}
  \label{eq:noisy_Hamiltonian}
  H(t) =
  \begin{cases}
    H_1, & t_n \leq t < t_n + T_{1n}, \\
    H_2, & t_n + T_{1n} \leq t < t_n + T_{1n} + T_{2n}.
  \end{cases}
\end{equation}
In this expression, $t_n$ is the elapsed time after a particular realization of
$n$ noisy driving cycles. The value of $t_n$ depends on all prior time shifts:
\begin{equation}
  \label{eq:t_n}
  t_n = \sum_{n' = 1}^n \sum_{i = 1, 2} T_{in'} = n T + \sum_{n' = 1}^n \sum_{i
    = 1, 2} \tau_{in'}. 
\end{equation}
Evidently, $t_n$ is itself a random variable. It's mean
$\overline{t_n} = n T + \sum_{n' = 1}^n \sum_{i = 1,2} \overline{\tau_{in}} = n
T$ coincides with the time span corresponding to $n$ noiseless driving periods,
while the fluctuations of $t_n$ grow as $\sqrt{n}$. The state of the system
after $n$ driving cycles is given by
\begin{equation}  
    \ket{\psi_n} = U_{ {\rm F}, n} \dotsb U_{ {\rm F}, 1} \ket{\psi_0}.
\end{equation}
Here, $U_{ {\rm F}, n}$ denotes a ``noisy'' Floquet operator. We obtain it from a
straightforward generalization of Eq.~\eqref{eq:UF_binary},
\begin{equation}
  \label{eq:UF_binary_noisy}
  U_{ {\rm F}, n} = \mathsf{T} e^{-i \int_{t_{n - 1}}^{t_n} dt \, H(t)} = e^{-i T_{2n} H_2} e^{-i T_{1n} H_1}.
\end{equation}

In full analogy, we can extend the kicking protocol defined by
Eq.~\eqref{eq:H_kicked} to include timing noise~\cite{Oskay2003, Bitter2016,
  Bitter2017, Cadez2017, Cadez2018}:
\begin{equation}
  \label{eq:H_kicked_noise}
  H(t) = H_1 + \lambda \sum_n \delta(t - t_n) H_2.
\end{equation}
With $t_n$ defined as in Eq.~\eqref{eq:t_n}, the waiting time between two
consecutive kicks is given by $t_{n + 1} - t_n = T + \tau_{1n}$. The resulting
noisy kicking protocol is described by a Floquet operator that takes exactly the
same form as the one for the binary Floquet system in
Eq.~\eqref{eq:UF_binary_noisy} if in addition to timing noise we allow for
fluctuations of the kicking strength, i.e., we replace
$\lambda \to \lambda + \tau_{2 n}$.

We have thus specified the noisy Floquet operator that determines the evolution
of the system's state $\ket{\psi_n}$ during one particular realization of a
noisy driving cycle as given in Eq.~\eqref{eq:psi_evolution}. We now proceed to
evaluate the average in Eq.~\eqref{eq:caligraphic-F} to obtain a formal
expression for the Floquet superoperator $\mathcal{F}$.

\subsection{Formal Solution}
\label{sec:formal-solution}

For a given distribution of the fluctuating times $T_{in}$ in
Eq.~\eqref{eq:UF_binary_noisy}, the average in Eq.~\eqref{eq:caligraphic-F} can
be carried out explicitly. To begin with, we focus on the first step of the
$n$th driving cycle. The time evolution of the density matrix during this step
is given by 
\begin{equation}
  \label{eq:strong_step_1}
  \rho_{n + 1, 1} = \overline{U_{n + 1, 1} \rho_n U_{n + 1, 1}^{\dagger}},
\end{equation}
where $U_{n + 1, 1} = e^{-i T_{1n} H_1}$. In order to perform the noise average,
we rewrite the time evolution in terms of the superoperator $\mathcal{H}_1$,
which is defined by $\mathcal{H}_1 A = [H_1, A]$ for a (matrix) operator
$A$. Equation~\eqref{eq:strong_step_1} can now be recast as
\begin{equation}
  \label{eq:super-strong_step_1}
  \rho_{n + 1, 1} = \overline{e^{-i T_{1n} \mathcal{H}_1}} \rho_n.
\end{equation}
A way to see the equivalence of Eqs.~\eqref{eq:strong_step_1}
and~\eqref{eq:super-strong_step_1} which proves useful in the following, is to
note that as is the case for operators $A$ acting on pure states $\ket{\psi}$,
functions of superoperators such as $\mathcal{H}_1$ can be written in terms of
their spectral representation: From an eigenbasis $\ket{\alpha}$ of $H_1$ with
$H_1 \ket{\alpha} = E_{\alpha} \ket{\alpha}$ we obtain the eigenoperators
$\ket{\alpha} \bra{\beta}$ of $\mathcal{H}_1$ which obey
$\mathcal{H}_1(\ket{\alpha} \bra{\beta}) = \left( E_{\alpha} - E_{\beta} \right)
\ket{\alpha} \bra{\beta}$.
Thus, the spectral representation of $\mathcal{H}_1$ reads
\begin{equation}
  \label{eq:spectral_H_1}
  \mathcal{H}_1 = \sum_{\alpha, \beta} \left( E_{\alpha} - E_{\beta} \right) \mathcal{P}_{\alpha \beta}, 
\end{equation}
with the superoperator $\mathcal{P}_{\alpha \beta}$ defined by the relation
$\mathcal{P}_{\alpha \beta} A = \ket{\alpha} \bra{\beta} \braket{\alpha | A | \beta}$. Analogously we
obtain
\begin{equation}
  e^{-i T_{1n} \mathcal{H}_1} \rho_n = \sum_{\alpha, \beta} e^{-i T_{1n} \left(
      E_{\alpha} - E_{\beta} \right)} \ket{\alpha} \bra{\beta} \braket{\alpha |
    \rho_n | \beta},
\end{equation}
see Eq.~\eqref{eq:UF}.
This expression can be seen to be equal to $U_{n + 1, 1} \rho_n U_{n + 1, 1}^{\dagger}$ by
inserting the completeness relation of the states $\ket{\alpha}$ twice in the
latter expression. While this shows how calculations with superoperators can be
carried out analytically, the concrete implementation of the superoperator
formalism for numerical purposes is discussed in Appendix~\ref{app:super}.

Returning to the average in Eq.~\eqref{eq:super-strong_step_1}, we note that its
result depends on the distribution of the duration $T_{1n}$. For noisy Floquet
systems, the average factorizes as
$\overline{e^{-i T_{1n} \mathcal{H}_1}} = e^{-i T_1 \mathcal{H}_1}
\overline{e^{-i \tau_{n1} \mathcal{H}_1}}.$
Exemplary distributions for timing noise are a normal distribution with width
$\tau$ and a uniform distribution on the interval
$[- \sqrt{3} \tau, \sqrt{3} \tau]$, both leading to fluctuations
$\overline{\tau_{n1}^2} = \tau^2$. We thus obtain
\begin{equation}
  \label{eq:avg_normal_uniform}
  \mathcal{E}_1 = \overline{e^{-i \tau_{n1} \mathcal{H}_1}} =
  \begin{cases}
    e^{- \tau^2 \mathcal{H}_1^2/2}, & \text{normal,} \\
    \sinc(\sqrt{3} \tau \mathcal{H}_1), & \text{uniform,}
  \end{cases}
\end{equation}
where $\sinc(x) = \sin(x)/x$ and we set $\sinc(0) = 1$. Using the spectral
representation Eq.~\eqref{eq:spectral_H_1} for the example of a uniform
distribution of timing errors we thus obtain
\begin{equation}
  \mathcal{E}_1 = \sum_{\alpha, \beta} \sinc(\sqrt{3 }\tau \left( E_{\alpha} -
    E_{\beta} \right)) \mathcal{P}_{\alpha \beta}.
\end{equation}
For any distribution of timing errors the full evolution during the first step
can be written as
\begin{equation}
  \label{eq:step_1_final}
  \rho_{n + 1, 1} = e^{-i T_1 \mathcal{H}_1} \mathcal{E}_1 \rho_n = \mathcal{U}_1
  \mathcal{E}_1\rho_n.
\end{equation}
Repeating the above reasoning for the second step of the time evolution leads us
to
\begin{equation}
  \label{eq:formal_twostep}
  \rho_{n+1} = \mathcal{F} \rho_n = \mathcal{U}_{2} \mathcal{E}_2 \mathcal{U}_1
  \mathcal{E}_1 \rho_n,
\end{equation}
where $\mathcal{E}_2$ is defined analogously to $\mathcal{E}_1$ in
Eq.~\eqref{eq:avg_normal_uniform}. As a consistency check, we note that for
vanishing timing noise, i.e., $\tau \to 0$, from
Eq.~\eqref{eq:avg_normal_uniform} we see that $\mathcal{E}_i = \id$, and thus
$\rho_{n+1} = \mathcal{U}_{\rm F} \rho_n$, where
$\mathcal{U}_{\rm F} = \mathcal{U}_2 \mathcal{U}_1 = e^{-iT_2\mathcal{H}_2} e^{- i T_1
  \mathcal{H}_1}$.
We now proceed to discuss the extension of the above derivation to multi-step
piecewise constant driving protocols.

\subsection{Extension to multi-step piecewise constant driving}
\label{sec:multi-phase-pc}

Extended driving protocols, which are defined in terms of a sequence of
Hamiltonians $H_i$ with $i = 1, 2, \dotsc, M$, can be treated in much the same
way as the binary driving of the previous section. To be specific, if
fluctuations of each of the $T_{in}$ are independent and obey the same
statistics, we find
\begin{equation}
  \label{eq:F-multistep}
  \rho_{n + 1} = \mathcal{F} \rho_n = \mathcal{U}_M \mathcal{E}_M \dotsb \mathcal{U}_1 \mathcal{E}_1
  \rho_n,
\end{equation}
where $\mathcal{U}_i = e^{-i T_i \mathcal{H}_i}$, and the $\mathcal{E}_i$ are
defined as in Eq.~\eqref{eq:avg_normal_uniform}. From this form it is
straightforward to obtain a matrix representation of $\mathcal{F}$ by using the
spectral representation of the superoperators $\mathcal{H}_i$ given in
Eq.~\eqref{eq:spectral_H_1}. Such a matrix representation can be used to find
the eigenmodes and complex quasienergies of $\mathcal{F}$ numerically.

Alternatively, we can iterate the procedure of Sec.~\ref{sec:formal-solution}
with one modification: A clear separation between the noiseless Floquet dynamics
and the noise-induced dissipation can be established by commuting
$\mathcal{U}_1$ in Eq.~\eqref{eq:formal_twostep} through the noisy part of
$\mathcal{E}_2$ of the second step. This can be done most easily before carrying
out the noise average and by using the relation (here, again, for a binary
driving protocol)
\begin{equation}
  \begin{split}
      U_{ {\rm F}, n} & = U_2 e^{-i \tau_{2n} H_2} U_1 e^{-i \tau_{1n} H_1} \\ & = U_{\rm F}
    e^{-i \tau_{2n} U_1^{\dagger} H_2 U_1} e^{-i \tau_{1n} H_1}.
  \end{split}
\end{equation}
Extending this procedure to a driving protocol that consists of $M$ steps, we
obtain
\begin{equation}
  \label{eq:F-alternative}
  \rho_{n + 1} = \mathcal{F} \rho_n = \mathcal{U}_{\rm F} \tilde{\mathcal{E}}_M \dotsb
  \tilde{\mathcal{E}}_1 \rho_n,
\end{equation}
where now
\begin{equation}
  \label{eq:E_i}
  \tilde{\mathcal{E}}_i =
  \begin{cases}
    e^{- \tau^2 \mathcal{L}_i^2/2}, & \text{normal,} \\
    \sinc(\sqrt{3} \tau \mathcal{L}_i), & \text{uniform.}
  \end{cases}
\end{equation}
The superoperators $\mathcal{L}_i$ act on operators $A$ according to
$\mathcal{L}_i A = [L_i, A]$, and the Hermitian operators $L_i$ are defined as
\begin{equation}
  \label{eq:L}  
  L_i = U_1^{\dagger} \dotsb U_{i - 1}^{\dagger} H_i U_{i - 1} \dotsb U_1.
\end{equation}
The expression Eq.~\eqref{eq:F-alternative} for the Floquet superoperator is most convenient for
studying a limiting case that is particularly relevant for experimental
realizations of Floquet systems: weak timing noise.

\subsection{Weak Noise: Discrete-Time Floquet-Lindblad Equation}
\label{sec:weak-noise-FLE}

As discussed in Sec.~\ref{sec:timing-noise}, timing noise occurs in Floquet
systems as a result of experimental imperfections. Therefore, it is often
justified to treat noise as a weak perturbation. In particular, if the spectra
of the Hamiltonians $H_i$ constituting the driving protocol are bounded by an
energy scale $J$ (e.g., the single-particle bandwidth in a tight-binding model)
which satisfies $\tau J \ll 1$, the error superoperators in Eq.~\eqref{eq:E_i}
can be expanded as
\begin{equation}
  \label{eq:strong_weak_expansion}  
  \tilde{\mathcal{E}}_i = \id - \frac{\tau^2}{2} \mathcal{L}_i^2. 
\end{equation}
We note that both normal and uniform distributions of time shifts lead to the
same lowest-order expansion. Inserting this expansion in Eq.~\eqref{eq:F-alternative} and keeping only terms up to
$O(\tau^2)$, we recover the FLE of Ref.~\cite{Rieder2017}, namely
\begin{equation}
  \label{eq:FLE}
  \rho_{n + 1} = \mathcal{U}_{\rm F} \left( \id + \tau^2 \sum_{i = 1}^M
    \mathcal{D}[L_i] \right) \rho_n.
\end{equation}
Here -- in reminiscence of the usual Markovian master equation in Lindblad
form -- we introduced the ``dissipator''
$\mathcal{D}[L_i] = - \mathcal{L}_i^2/2$ for ``jump operators'' $L_i$. The similarity becomes obvious when 
writing the double-commutator structure of Eq.~\eqref{eq:FLE} for Hermitian jump
operators $L = L^{\dagger}$, namely
\begin{equation}
  \label{eq:dissipator}
  - \frac{1}{2}
  \mathcal{L}^2 \rho= - \frac{1}{2} \left[ L, \left[ L, \rho \right] \right] 
  =L \rho L - \frac{1}{2} \left\{ L^2, \rho \right\} =
   \mathcal{D}[L] \rho 
\end{equation}
where $\mathcal{L} \rho = [L, \rho]$. Markovian master equations in Lindblad
form find widespread use in quantum optics, where the time-local form of the
equation results from a separation of scales between the typical time scale of
the system's evolution and the much shorter coherence time of the bath which
induces dissipation in the system. Similarly, the FLE~\eqref{eq:FLE} is local in
driving cycles, i.e., the state of the system $\rho_{n + 1}$ after $n + 1$
cycles depends only on the state $\rho_n$ after $n$ cycles and not on
$\rho_{n'}$ with $n' < n$. This is due to our assumption that fluctuations of
the stochastic drive are uncorrelated between different driving cycles, which is
analogous to the Markovian baths encountered in the quantum optics
context~\cite{Reimer2018}. Finally, we note that while here we found the FLE
equation in the limit of weak noise, it is not guaranteed that the generator of
the Floquet superoperator is of Lindblad form~\cite{Schnell2018}.

\section{Fully random driving}
\label{sec:fully-random-driving}

Having given a detailed account of Floquet systems with timing noise, we now
turn to a physical situation whose relation to any form of periodicity is
perhaps less obvious: a time-dependent system subject to random telegraph noise
(RTN)~\cite{PhysRevA.24.398, HuYing2015, Cai2016}. We show how randomly timed
pulses can be treated as an extreme case within our Floquet superoperator
formalism.

\subsection{Random telegraph noise}
\label{sec:rand-telegr-noise}

For $\tau \to 0$, the evolution of the pure state of the noisy Floquet
system Eq.~\eqref{eq:psi_evolution} considered in the previous section reduces to
the familiar perfectly periodic Floquet form. Here, instead, we consider a
system that is driven exclusively by noise. As an example, consider a
time-dependent Hamiltonian $H(t) = H_0 + \lambda(t) V$, in which the random
process $\lambda(t)$ jumps between two values $\lambda_{1,2}$ (thus, $H(t)$
jumps between $H_{1,2} = H_0 + \lambda_{1,2} V$) at a rate $\gamma$. Under these
conditions, the waiting times between sudden quenches from $H_1$ to $H_2$ and
back are indeed statistically independent and follow an exponential distribution
\begin{equation}
  \label{eq:exp_distribution}
  P(t) = \gamma e^{- \gamma t}.
\end{equation}
The average duration of a driving cycle
corresponding to starting with $H_1$ and waiting for the Hamiltonian to change
to $H_2$ and back is thus $T = 2/\gamma$. As above, the noise averaged evolution
of the system's density matrix is given by Eq.~\eqref{eq:caligraphic-F} with
the noisy Floquet operator defined in Eq.~\eqref{eq:UF_binary_noisy}, but with
$T_{in}$ drawn from the same exponential distribution. However, in the present
case there is no meaningful definition of noise strength and thus of a noiseless
limit. In the limiting case $\gamma \to 0$ the Hamiltonian of the system is
simply $H(t) = H_1$ up to arbitrarily long times, while for $\gamma \to \infty$
corresponding to fast switching between $H_1$ and $H_2$ we show below 
that the time evolution follows the average
Hamiltonian $\overline{H} = (H_1 + H_2)/2$ with only slow dephasing.

The formal results of Sec.~\ref{sec:formal-solution} can be generalized to
this case of RTN by using the exponential distribution for the timing noise,
yielding now
\begin{equation}
  \label{eq:avg_exp}
  \mathcal{E}_1 = \overline{e^{-i \tau_{n1} \mathcal{H}_1}} = \frac{\gamma}{\gamma + i \mathcal{H}_1}.
\end{equation}
and, with $T_i=0$, the ``Floquet'' step reduces to
$\mathcal{U}_1 = e^{-i T_1 H_1} = \id$.
\subsection{Fast switching}
\label{sec:fast-switching}

The evolution of a system driven by RTN is given by [cf.\
Eqs.~\eqref{eq:formal_twostep} and~\eqref{eq:avg_exp}]
\begin{equation}
  \label{eq:RTN_evolution}
  \rho_{n + 1} = \frac{\gamma}{\gamma + i \mathcal{H}_2} \frac{\gamma}{\gamma +
    i \mathcal{H}_1} \rho_n.
\end{equation}
For fast switching rates, when $\gamma$ is much larger than the spectral
bandwidth of the Hamiltonians $H_i$, a leading-order expansion in $1/\gamma$
yields
\begin{equation}
  \label{eq:motional_narrowing}
  \rho_{n + 1} = \left[ \id - \frac{i}{\gamma} \left( \mathcal{H}_1 +
      \mathcal{H}_2 \right) \right] \rho_n = e^{- i T (\mathcal{H}_1 +
    \mathcal{H}_2)/2} \rho_n,
\end{equation}
where $T = 2/\gamma$ is the average period, and the exponentiation is valid up
to the same order in $1/\gamma$. Equation~\eqref{eq:motional_narrowing}
describes coherent evolution with the average Hamiltonian $(H_1 +
H_2)/2$. Noise-induced dissipation is suppressed at fast switching rates, and
occurs only at higher orders in $1/\gamma$. This suppression of decoherence at
high switching rates has also been found in a description of the RTN-driven
dynamics using a generalized master equation for the marginal system density
operator~\cite{PhysRevA.24.398, HuYing2015, Cai2016}.

\section{Application to noisy Floquet topological chains}
\label{sec:appl-noisy-floq}

We now turn to a concrete application of the formalism of statistical
periodicity in driven quantum systems. In our recent work~\cite{Rieder2017}, we
used a Floquet-Lindblad equation to describe the loss of an end state in a noisy
Floquet topological chain. Here, we re-derive these results using the Floquet
superoperator formalism introduced above.

\subsection{Model}
\label{sec:model}

\begin{figure}
  \centering
    \includegraphics[width=.45\textwidth]{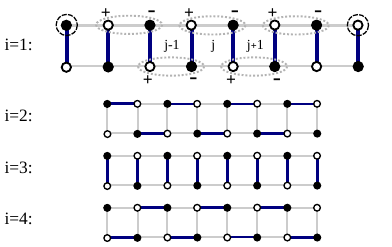}
    \caption{Sketch of the model considered in
      Sec.~\ref{sec:appl-noisy-floq}. Lattice sites form doublets (gray dashed
      ellipses) and are labeled with the plaquette number $j$ and a doublet
      index $s=\pm$. The driving protocol comprises four steps $i=1,2,3,4$
      during which hopping along only certain ``active'' bonds is allowed, as
      indicated with the thick blue lines. As a result of the particular
      driving, localized states, indicated by dashed black circles, form at the
      ends of the ladder at lattice sites $(0,-)$ and $(L,+)$.}
  \label{fig:floquet_1D}
\end{figure}

We consider a periodically time-dependent system of non-interacting spinless
fermions on a one-dimensional ladder, implemented by varying the hopping
amplitudes between neighboring lattice sites. One period comprises four steps of
equal duration $ T/4$ during each of which the system parameters are held
constant. The switches between the different phases happen instantaneously. This
is described by the Hamiltonian
\begin{equation}
  \label{eq:Hamiltonian-Floquet-topo-chain}
  H(t) = H_i \quad \text{for} \quad \left( i-1 \right) T/4 \leq t < iT/4,
\end{equation}
where the time is measured modulo the period $T$ and
\begin{align}
  \label{eq:Hamiltonian-step-Floquet-topo-chain}
  H_i = - \sum_{\mu, \nu} J_{\mu \nu}^i \left( c_{\mu}^{\dagger} c^{\phantom{\dag}}_{\nu} + \Hc \right),
\end{align}
for steps $i=1,2,3,4$. The sum runs over a combined index for sites on the
ladder, $\mu = (j,s)$, with $j=0, \dotsc, L$ labeling doublets of sites at the
plaquettes of the ladder and $s=\pm$ denoting their sublattice index as
illustrated in Fig.~\ref{fig:floquet_1D}. The hopping amplitudes
$ J_{\mu \nu}^i $ are chosen to enable hopping along disconnected pairs of
nearest-neighbor lattice sites. In particular, we set $ J_{\mu \nu}^i = J$ for
all active bonds $\{\mu,\nu\}_i$ of step $i$ and $J_{\mu \nu}^i = 0 $
otherwise. We define the driving protocol by setting the active bonds as
\begin{equation}
  \begin{split}
    \{\mu,\nu\}_1 &= \{(j,s), (j-s,-s)\} \\
    \{\mu,\nu\}_2 &= \{(j,s), (j-2s,-s)\} \\
    \{\mu,\nu\}_3 &= \{(\mu,\nu)\}_1 \\
    \{\mu,\nu\}_4 &= \{(j,s), (j,-s) \}.
  \end{split}
\end{equation}

At a special point in parameter space $J = 2 \pi/T$, which we call ``resonant
driving,'' the eigenstates of the Floquet operator corresponding to the
time-dependent Hamiltonian Eq.~\eqref{eq:Hamiltonian-Floquet-topo-chain} can be
understood intuitively: Each step of the driving protocol (of duration $T/4$)
results in the full transfer of particles between two coupled neighboring sites,
such that each particle accumulates a phase $\pi/2$. Therefore, during one
period, a particle which is initialized on a single lattice site performs a full
circle around a plaquette, thereby collecting a phase factor $2\pi$. The only
exceptions to this behavior are particles initialized on one of the two lattice
sites at the ends of the chain as indicated in Fig.~\ref{fig:floquet_1D}. These
skip two steps of the driving protocol, returning back to their original
positions with a phase of $\pi$. We thus find a completely flat bulk band at
quasienergy $0$ and two topologically protected~\cite{Rieder2017} end states at
quasienergy $\varepsilon T = \pi$. For a chain of length $L$, the left and right
end states are $\ket{e_l} = \ket{0, -}$ and $\ket{e_r} = \ket{L, +}$.  When we
tune the system away from the point of resonant driving, the bulk band becomes
dispersive, while the end states remain at quasienergy $\varepsilon T = \pi$,
acquiring only a finite localization length.

\begin{figure}
 \includegraphics[width=\columnwidth]{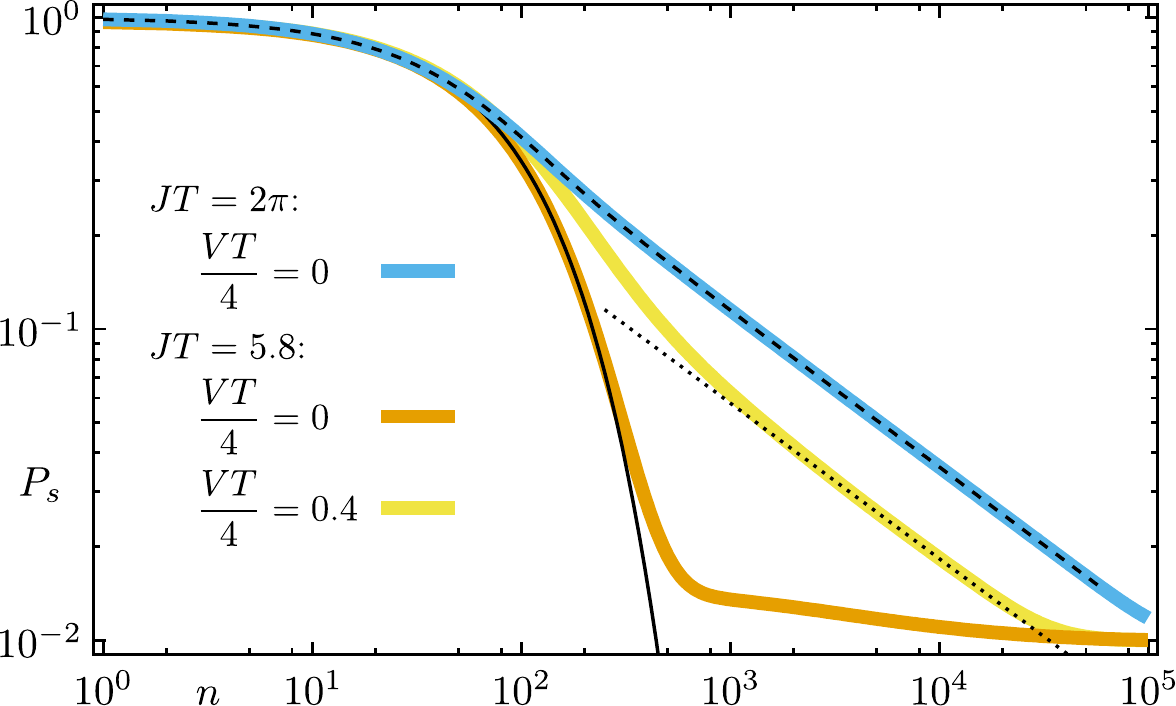}
 \caption{Survival probability of the end mode computed using
   Eq.~\eqref{eq:survival_superoperator} as a function of the number of noisy
   driving cycles, $n$. We use a ladder consisting of 50 rungs and a noise
   strength $\tau/T=1/80$. Away from the resonant driving point ($JT=5.8$,
   orange curve), $P_{\mathrm{s}}$ decays exponentially with $n$. The decay
   slows down to a diffusive one either when fine-tuning the system to $JT=2\pi$
   (blue), or by including disorder ($VT/4=0.4$, yellow). The disordered curve
   is obtained by averaging over 80 simultaneous noise-disorder realizations,
   with error bars smaller than the line width. The solid black line shows an
   exponential decay with rate $2J^2\tau^2\simeq0.0105$ as predicted in
   Eq.~\eqref{eq:survival-exponential}, the dotted line is a fit to a diffusive
   decay $\sim t^{-1/2}$, whereas the dashed line is the analytic result of
   Eq.~\eqref{eq:survival-bessel}.\label{fig:survivals}}
\end{figure}

The observation of these localized end states in an experimental realization
would provide a clear signature of the non-trivial topological properties of the
model. In Ref.~\cite{Rieder2017}, we studied how the observability of end states
is affected by timing noise of the type described in
Sec.~\ref{sec:noisy-floq-syst}. Timing noise causes a particle which is
initialized in an end state to decay into the system's bulk. We showed that the
nature of this decay depends critically on the nature of the bulk: For a
delocalized bulk, noise-induced excitations out of the end state propagate away
freely, resulting in an exponential decay with time. In contrast, when the bulk
is localized these excitations ``get stuck'' and have a finite probability of
returning to the end state. This results in a dramatically slowed-down diffusive
decay. The different behaviors are shown in Fig.~\ref{fig:survivals}.

To obtain localized bulk states in the Floquet ladder model, it is sufficient to
consider the resonant driving point, $J=2\pi/T$, at which bulk states are
dispersionless. In an experimental setting however, fine-tuning the system
parameters to this point may be impractical, and a more accessible means of
localizing bulk states is by adding quenched (time-independent) disorder to the
system. To model the latter scenario, we consider random on-site disorder, adding
to the Hamiltonian a term of the form
\begin{equation}\label{eq:Hdis}
    H_{\rm dis} = \sum_\mu V_\mu c^\dag_\mu c^{\phantom{\dag}}_\mu,
\end{equation}
where the potential $V_\mu$ is drawn randomly and independently for each lattice
site of the ladder from the uniform distribution $[-V/2,V/2]$, with $V$ the
disorder strength.

Note that on-site disorder destroys the topological
protection of the boundary modes, since in the presence of a non-zero chemical
potential they are allowed to shift away from the quasienergy zone boundary and
hybridize with the bulk states. However, as shown in Ref.~\cite{Rieder2017}, for
small values of the disorder strength $V$, end modes are still well separated
from bulk states and localized at the boundaries of the system, such that
studying their decay is a well defined problem.

In the following, we re-derive the results on the end state decay presented in
Ref.~\cite{Rieder2017} using the concepts and tools developed in the present
paper.

\subsection{Decay of the end state}
\label{sec:decay-end-state}

We consider the stroboscopic time evolution of a particle that is initialized in
the left end state $\ket{e} = \ket{e_l}$ of the chain, i.e., the initial density
matrix is given by $\rho_0 = \ket{e} \bra{e}$. The quantity of interest is the
survival probability $P_{\mathrm{s}}$ of the end state which is defined as
\begin{equation}
  \label{eq:survival-1}
  P_{\mathrm{s}} = \overline{\abs{\braket{e | \psi_n}}^2} = \bra{e}
  \overline{\ket{\psi_n} \bra{\psi_n}} \ket{e} = \braket{e | \rho_n | e},
\end{equation}
and is the probability to find a particle in the end state after $n$ driving
cycles. In the absence of noise, $P_{\mathrm{s}} = 1$ stays constant, while
$P_{\mathrm{s}}$ decays over time for noisy driving. The dynamics of the
system's density is determined by the evolution Eq.~\eqref{eq:caligraphic-F},
i.e., the time-evolved state is $\rho_n = \mathcal{F}^n \rho_0$. Before we
specify the Floquet superoperator $\mathcal{F}$ for the noisy Floquet
topological chain, we show how the survival probability can be evaluated in the
superoperator formalism.

\subsubsection{Survival probability from the Floquet superoperator}
\label{sec:surv-prob-F}

In the following, we find it convenient to adopt the terminology and notation of
Ref.~\cite{Zwolak2004}. That is, we regard operators as ``superkets'' which we
distinguish from normal kets (i.e., state vectors) by a subscript $\sharp$. To
emphasize this interpretation, we write the density matrix as
$\rho_n = \kket{\rho_n}$ and $\ket{e} \bra{e} = \kket{e}$ for the projector on
the (left) end state of the topological chain to name two examples. A scalar product of superkets is
given by $\bbraket{A | B} = \tr(A^{\dagger} B)$. Thence, the survival
probability of the end state Eq.~\eqref{eq:survival-1} can be written as an
overlap of superkets:
\begin{equation}
  \label{eq:survival-2}
  P_{\mathrm{s}} = \tr( \ket{e} \bra{e}
  \overline{\ket{\psi_n} \bra{\psi_n}} ) = \bbraket{e | \rho_n},
\end{equation}
where according to Eq.~\eqref{eq:caligraphic-F} the time-evolved density-matrix
superket reads $\kket{\rho_n} = \mathcal{F}^n \kket{\rho_0}$. As in unitary
Hamiltonian dynamics, a convenient representation of the time-evolved state and
thus of the survival probability Eq.~\eqref{eq:survival-2} can be given by expanding
$\kket{\rho_n}$ in a basis of eigenoperators of $\mathcal{F}$. In particular, we
denote the right eigenoperators of $\mathcal{F}$ by $\kket{\alpha}$,
\begin{equation}\label{eq:FSO_evals}
  \mathcal{F} \kket{\alpha} = e^{-i \lambda_{\alpha}} \kket{\alpha}.
\end{equation}
Since, in general, the superoperator $\mathcal{F}$ is not normal, we have to
distinguish its left and right eigenoperators. Denoting by $\bbra{\alpha}$ the
left eigenoperator corresponding to the eigenvalue $\lambda_{\alpha}$, the
identity superoperator can be written as
$\id = \sum_{\alpha} \kket{\alpha} \! \bra{\alpha}$ (we use the symbol $\id$
both for the identity operator and the identity superoperator).  The
stroboscopic time evolution can thus be written as
\begin{equation}
  \kket{\rho_n} = \mathcal{F}^n \kket{\rho_0} = \sum_{\alpha} e^{-i n \lambda_{\alpha}}
  \bbraket{\alpha | \rho_0} \kket{\alpha}.
\end{equation}
Inserting this representation with $\kket{\rho_0} = \kket{e}$ in
Eq.~\eqref{eq:survival-2}, the survival probability of the end state becomes
\begin{equation}
  \label{eq:survival_superoperator}
  P_{\mathrm{s}} = \sum_{\alpha} e^{-i n \lambda_{\alpha}} \abs{\bbraket{\alpha | e}}^2.
\end{equation}
$P_{\mathrm{s}}$ is evidently fully determined by the eigenoperators and eigenvalues of the
Floquet superoperator $\mathcal{F}$. We proceed by specifying $\mathcal{F}$ for
the noisy Floquet topological chain, and then evaluate the survival
probability Eq.~\eqref{eq:survival_superoperator} for dispersive and localized
bulk states, leading to exponential and diffusive decay, respectively, in
Secs.~\ref{sec:expon-decay-disp},~\ref{sec:diff-decay-reson},
and~\ref{sec:diff-decay-disord}.

\subsubsection{Floquet superoperator for a noisy Floquet topological chain}
\label{sec:floq-super-noisy}

The formal expression for the Floquet superoperator for the Floquet topological
chain with timing noise is given by Eq.~\eqref{eq:F-multistep}, where the
coherent parts of the evolution, $\mathcal{U}_i = e^{-i T_i \mathcal{H}_i}$, are
generated by the Hamiltonians $H_i$ in
Eq.~\eqref{eq:Hamiltonian-step-Floquet-topo-chain} (recall that $\mathcal{H}_i$
is defined by its action on an operator $A$, which is
$\mathcal{H}_i A = [H_i, A]$), and the error operators for normally distributed
timing noise take the form given in Eq.~\eqref{eq:avg_normal_uniform}, that
is, the noise-averaged Floquet superoperator reads
\begin{equation}
  \label{eq:F-noisy-Floquet-topo-chain}
  \mathcal{F} = \mathcal{U}_4 \mathcal{E}_4 \dotsb \mathcal{U}_1
  \mathcal{E}_1 = e^{- \frac{i T}{4} \mathcal{H}_4} e^{-\frac{\tau^2}{2}
    \mathcal{H}_4^2} \dotsb e^{- \frac{i T}{4} \mathcal{H}_1}
  e^{-\frac{\tau^2}{2} \mathcal{H}_1^2}.
\end{equation}
To calculate the survival probability of the end state given in
Eq.~\eqref{eq:survival_superoperator} we have to find the spectrum and the
eigenoperators of $\mathcal{F}$. This can be done numerically as described
further below, but close to resonant driving we can also make progress
analytically. For this purpose, it is more convenient to work with the
alternative representation of the Floquet superoperator given in
Eq.~\eqref{eq:F-alternative}, which in the present case becomes
\begin{equation}
  \label{eq:F-alternative-noisy-Floquet-topo-chain}
  \mathcal{F} = \mathcal{U}_{\rm F} e^{- \frac{\tau^2}{2} \mathcal{L}_4^2} e^{-
    \frac{\tau^2}{2} \mathcal{L}_3^2} e^{- \frac{\tau^2}{2} \mathcal{L}_2^2}
  e^{- \frac{\tau^2}{2} \mathcal{L}_1^2}.
\end{equation}
The jump superoperators $\mathcal{L}_i$ are defined as $\mathcal{L}_i A = [L_i,
A]$, with the operators $L_i$ given in Eq.~\eqref{eq:L}. At resonant driving,
the latter take the form~\cite{Rieder2017}
\begin{equation}
  \label{eq:L_i-resonant}
  \begin{split}
    L_1 & = J \sum_{j} \left( \ket{j,+}\bra{j-1,-} + \Hc \right), \\
    L_2 & = J \sum_{j} \left( \ket{j,+}\bra{j,-} + \Hc \right), \\
    L_3 & = J \sum_{j} \left( \ket{j,+}\bra{j+1,-} + \Hc \right), \\
    L_4 & = J \sum_{j} \left( \ket{j,+}\bra{j,-} + \Hc \right).
  \end{split}
\end{equation}
This is the starting point of our analytical calculation of the end state's
decay close to resonant driving presented in the following. To extend the
analysis beyond this limiting case, but also to include disorder in the model,
it is necessary to determine the spectrum of the Floquet superoperator
numerically. A matrix representation of $\mathcal{F}$ that is amenable to a
numerical calculation of its eigenoperators and eigenvalues can be obtained by
introducing a basis in the space of operators as described in
App.~\ref{app:super}. All numerical results shown below are based on this
representation.

\subsubsection{Exponential decay for a dispersive bulk}
\label{sec:expon-decay-disp}

We first consider the case of a dispersive bulk, in which the survival
probability of the end state decays exponentially. Here, we show this
analytically for a system that is close to but crucially slightly away from
resonant driving, i.e., with $JT/4 = \pi/2 + \delta \phi$ where
$\delta \phi \ll 1$. Then, the projector on the end state,
$\kket{e} = \ket{e} \bra{e}$, is an approximate eigenstate of the Floquet
superoperator with $\lambda_e = -i 2 \kappa^2$ where $\kappa = J \tau$, which
according to Eq.~\eqref{eq:survival_superoperator} immediately implies the
asserted exponential decay.

To obtain these results, we work in a basis of eigenoperators of the noiseless
Floquet operator $\mathcal{U}_{\rm F}$. Starting from a basis of the Floquet operator
$U_{\rm F}$ which consists of bulk states $\ket{b}$ and the left end state $\ket{e}$
(we disregard the right end state assuming that the system size is much larger
than the localization length of the end states), a basis of $\mathcal{U}_{\rm F}$ is
formed by the superkets $\kket{b; b'} = \ket{b} \bra{b'}$,
$\kket{b; e} = \ket{b} \bra{e}$ (and its Hermitian conjugate), and
$\kket{e} = \ket{e} \bra{e}$. To see that $\kket{e}$ is an approximate
eigenoperator of $\mathcal{F}$ given in
Eq.~\eqref{eq:F-alternative-noisy-Floquet-topo-chain} it is sufficient to show
that the matrix elements $\bbraket{b; b' | \mathcal{L}_i^2 | e}$ and
$\bbraket{b; e| \mathcal{L}_i^2 | e}$ vanish in the thermodynamic limit. This,
together with $\mathcal{U}_{\rm F} \kket{e} = \kket{e}$, establishes the result. We
thus consider first the matrix elements
\begin{multline}
  \label{eq:bb-Li2-e-matrix-element}
  \bbraket{b; b' | \mathcal{L}_i^2 | e} \\
  \begin{aligned}
    & = \braket{b | \left( L_i^2 \ket{e} \bra{e} + \ket{e} \bra{e} L_i^2 - 2 L_i
        \ket{e} \bra{e} L_i \right) | b'} \\ & = -2 \braket{b | L_i | e}
    \braket{e | L_i | b'},
  \end{aligned}
\end{multline}
where in the second equality we used that $\braket{e | b} = 0$. In the above
relation, both the operators $L_i$ and the states $\ket{b}$ and $\ket{e}$ depend
on $J$. To find the leading behavior close to the resonant value
$JT=2\pi$ for the jump operators $L_i$ we can use the resonant form in
Eq.~\eqref{eq:L_i-resonant}. As with regard to the states, we note that for any
small deviation from resonant driving the bulk states are delocalized, and hence
$\braket{e | L_i | b} \sim J/\sqrt{L}$. Therefore, in the thermodynamic limit,
the matrix element in Eq.~\eqref{eq:bb-Li2-e-matrix-element} vanishes,
$\bbraket{b; b' | \mathcal{L}_i^2 | e} = 0$. To obtain the matrix elements
$\bbraket{b; e| \mathcal{L}_i^2 | e}$ and the diagonal element
$\bbraket{e | \mathcal{L}_i^2 | e}$ to leading order in $\delta \phi \to 0$, it
is helpful to note first that on resonance the diagonal elements of the jump
operators vanish, $\braket{e | L_i | e} = 0$, as can be seen immediately from
Eq.~\eqref{eq:L_i-resonant} and $\ket{e} = \ket{0, -}$. Second, the end state is
an eigenstate of $L_i^2$~\cite{Rieder2017},
\begin{equation}  
  \label{eq:left_edge_L2}
  L_i^2 \ket{e} =
  \begin{cases}
    J^2 \ket{e}, & i=1,3, \\
    0, & i = 2,4.
  \end{cases}
\end{equation}
Using these results, we see at once that the following off-diagonal matrix
elements vanish:
\begin{multline}
  \label{eq:be-Li2-e-matrix-element}
  \bbraket{b; e | \mathcal{L}_i^2 | e} \\
  \begin{aligned}
    & = \braket{b | \left( L_i^2 \ket{e} \bra{e} + \ket{e} \bra{e} L_i^2 - 2 L_i
        \ket{e} \bra{e} L_i \right) | e} \\ & = \braket{b | L_i^2 | e} - 2
    \braket{b | L_i | e} \braket{e | L_i | e},
  \end{aligned}
\end{multline}
while for the diagonal element we obtain
\begin{equation}
  \label{eq:e-Li2-e}  
  \bbraket{e | \mathcal{L}_i^2 | e} = 2 \left( \braket{e | L_i^2 | e} -
    \braket{e | L_i | e}^2 \right) =
  \begin{cases}
    2 J^2, & i = 1,3, \\
    0, & i = 2,4.
  \end{cases}
\end{equation}
Therefore, neglecting terms that vanish in the thermodynamic limit or are
$O(\delta \phi)$, the projector on the end state is an eigenoperator of the
squared superoperators $\mathcal{L}_i^2$, and thus of the Floquet
superoperator Eq.~\eqref{eq:F-alternative-noisy-Floquet-topo-chain}, 
\begin{equation}
  \label{eq:exp-decay}
  \mathcal{F} \kket{e} = e^{- 2 \kappa^2} \kket{e}.
\end{equation}
Inserting this in Eq.~\eqref{eq:survival_superoperator} yields exponential decay
of the survival probability,
\begin{equation}
  \label{eq:survival-exponential}
  P_{\mathrm{s}} = e^{- 2 n \kappa^2},
\end{equation}
with decay rate $2 \kappa^2 = 2 J^2 \tau^2$ as shown in
Fig.~\ref{fig:survivals}. In Ref.~\cite{Rieder2017}, we obtained the same result for
$\kappa \ll 1$ by considering the time evolution of $\kket{e}$ in a weak noise
expansion.

\begin{figure}
 \includegraphics[width=\columnwidth]{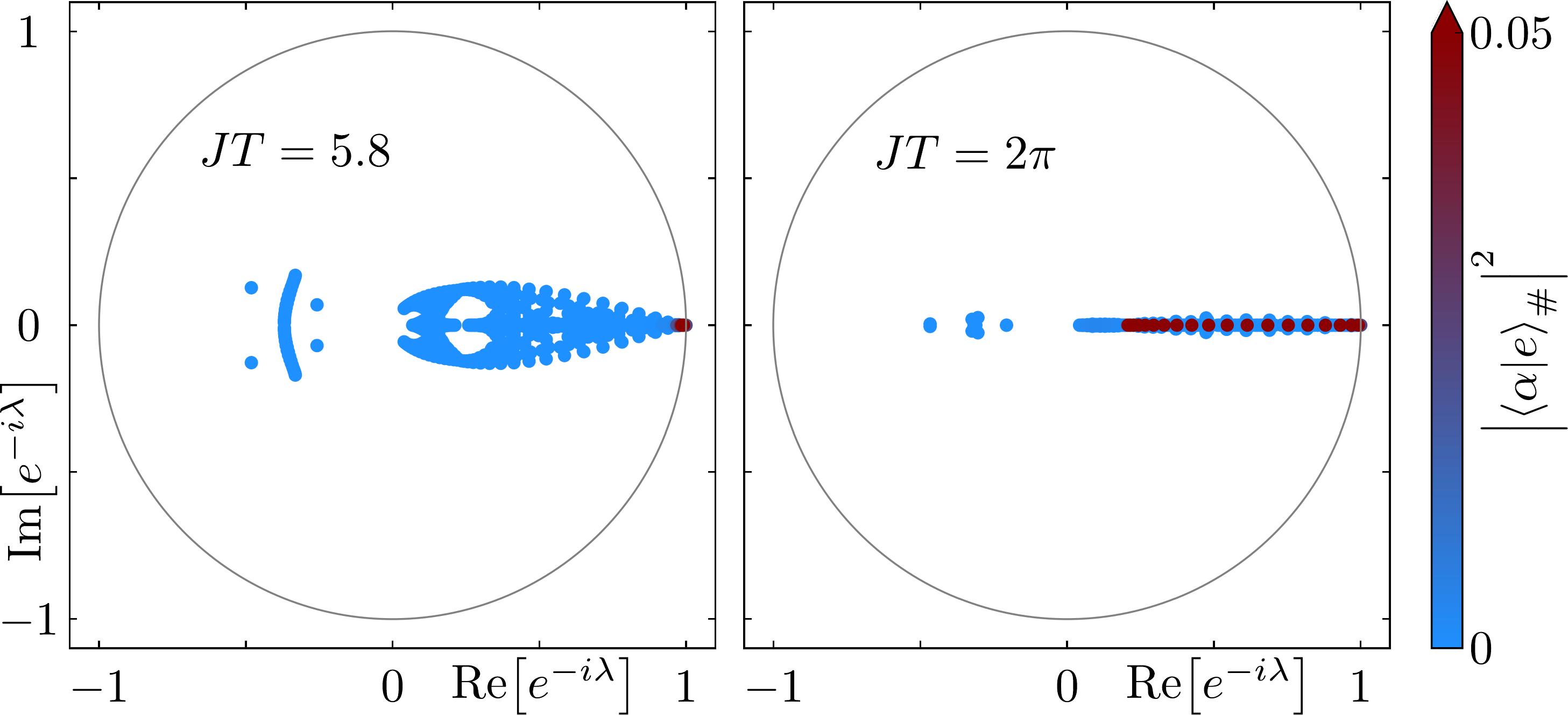}
 \caption{Real and imaginary parts of the eigenvalues of the Floquet
   superoperator ${\cal F}$ [Eq.~\eqref{eq:FSO_evals}], obtained numerically for
   a topological chain consisting of 20 rungs, with a noise strength
   $\tau/T=0.1$. Points inside the unit circle correspond to non-stationary
   states, which have negative imaginary parts, ${\rm Im}(\lambda)<0$, and the
   color scale denotes the squared overlap of each superket with the edge
   density matrix, $| \bbraket{\alpha|e}|^2$. At the resonant driving point
   ($JT=2\pi$, right panel) there is a larger number of states with a
   significant overlap than away from resonant driving ($JT=5.8$, left
   panel).\label{fig:Fevals_circle}}
\end{figure}

\begin{figure}
 \includegraphics[width=\columnwidth]{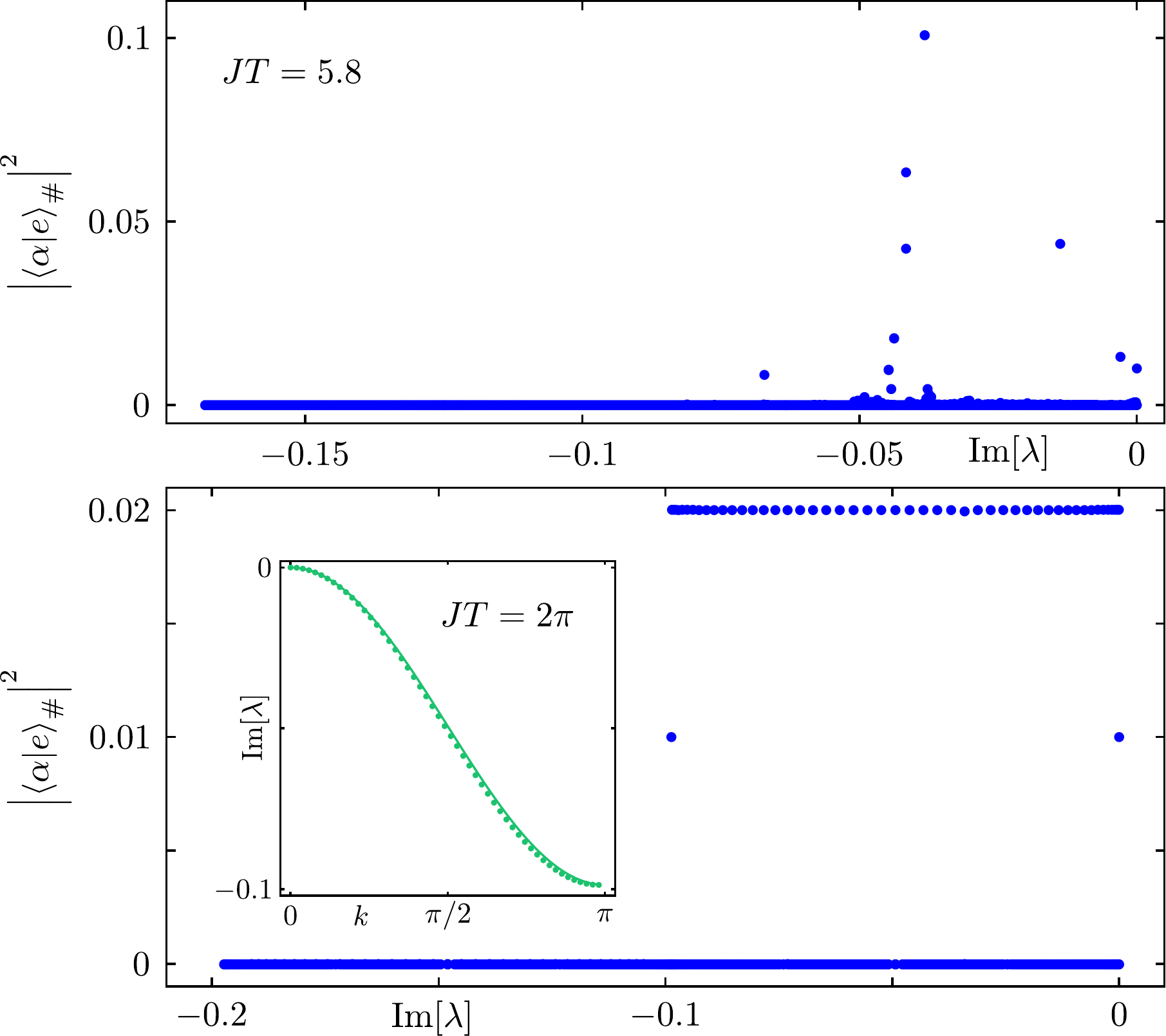}
 \caption{The squared overlap of the eigenoperators of ${\cal F}$ with the edge
   superket is plotted as a function of the imaginary part of the corresponding
   eigenvalues. Away from the resonant driving point (top panel), only a few
   eigenoperators have large values of $|\bbraket{\alpha|e}|^2$, leading to an
   exponential decay.  At resonant driving however (bottom panel), many states
   have a large and equal overlap with the edge, leading to a diffusive decay of
   the survival probability. We use a ladder of 50 rungs and $\tau/T=1/40$ in
   both panels. The inset contains only those eigenoperators for which
   $|\bbraket{\alpha|e}|^2>0.005$, using the data from the bottom panel. They
   are sorted in decreasing order of ${\rm Im}[\lambda]$ as
   $\lambda_0 > \lambda_1 > \ldots > \lambda_{N-1}$ and plotted as a function of
   $k=\pi j/N$, where $j\in\{ 0,N-1 \}$. For comparison, the solid curve shows
   the predicted behavior of $\lambda_\rho(k)$ from
   Eq.~\eqref{eq:lambda-rho-sigma-weak-noise}.\label{fig:Fevals_imag}}
\end{figure}

Figure~\ref{fig:Fevals_circle} shows the eigenvalues of the Floquet
superoperator ${\cal F}$ [Eq.~\eqref{eq:F-noisy-Floquet-topo-chain}] obtained numerically as well as
their squared overlaps with the edge superket, $\kket{e}$. As illustrated in the
left panel of the figure, away from the resonant driving point only few
eigenoperators have large values of $|\bbraket{\alpha|e}|^2$. This is consistent
with the above analytical result that the edge superket is an approximate
eigenstate of $\mathcal{F}$, and results in an exponential decay as per
Eq.~\eqref{eq:survival_superoperator}.  To better visualize this behavior, we
plot in Fig.~\ref{fig:Fevals_imag} the edge overlap as a function of the
imaginary part of the eigenvalues of ${\cal F}$.
Again, from the data shown in the top panel of the figure it is evident that
only few eigenoperators have a sizable overlap with the edge superket. Finally,
using the numerically obtained eigenoperators and eigenvalues of the Floquet
superoperator, we calculate the survival probability of the edge mode using
Eq.~\eqref{eq:survival_superoperator}, and find excellent agreement with the
analytical result Eq.~\eqref{eq:survival-exponential} as shown in
Fig.~\ref{fig:survivals}.

\subsubsection{Diffusive decay for resonant driving}
\label{sec:diff-decay-reson}

We now turn to exactly resonant driving, where the addition of timing noise
leads to diffusive decay of the end state survival probability. Here, we show
this explicitly by calculating the eigenoperators and corresponding eigenvalues
of the Floquet superoperator that have non-vanishing overlap with the end state
and thus contribute to the survival probability in
Eq.~\eqref{eq:survival_superoperator}.  This calculation is facilitated by the
simple form of the jump operators Eq.~\eqref{eq:L_i-resonant} on
resonance. Moreover, since the bulk band of the noiseless Floquet operator is
flat, we can work in the basis of lattice sites $\ket{j, s}$ and treat both end
and bulk states on the same footing. Thence, we consider for simplicity an
infinite chain, where $U_{\rm F} = \id$ and $\mathcal{U}_{\rm F} = \id$ is the identity
superoperator.

As we show below, the Floquet superoperator is block-diagonal in operator space,
and we diagonalize it in the subspace spanned by the projection operators on
single lattice sites, $\kket{j, s} = \ket{j, s} \bra{j,s}$. Indeed, using the
form of the jump operators given in Eq.~\eqref{eq:L_i-resonant}, it is
straightforward to check that
\begin{equation}  
  \begin{split}
    \mathcal{L}_1^2 \kket{j, s} & = 2 J^2 \left( \kket{j, s} - \kket{j - s, -s}
    \right), \\
    \mathcal{L}_3^2 \kket{j, s} & = 2 J^2 \left( \kket{j, s} - \kket{j + s, -s}
    \right), \\
    \mathcal{L}_2^2 \kket{j, s} & = \mathcal{L}_4^2 \kket{j, s} = 2 J^2 \left(
      \kket{j, s} - \kket{j, - s} \right).
  \end{split}
\end{equation}
The action of the $\mathcal{L}_i^2$ does not lead out of the spaced spanned by
the operators $\kket{j, s}$, i.e., $\mathcal{L}_i^2$ indeed assumes the asserted
block-diagonal form. Translational invariance suggests to diagonalize
$\mathcal{F}$ in a momentum space basis of superkets $\kket{k, s}$ defined by
\begin{equation}
  \label{eq:momentum-space-superkets}
  \kket{k, s} = \sum_j e^{-i k j} \kket{j, s}, \quad \kket{j, s} =
  \int_{-\pi}^{\pi} \frac{d k}{2 \pi} e^{i k j} \kket{k, s}.
\end{equation}
In this basis, the squared jump superoperators take the form
\begin{equation}
  \begin{split}
    \mathcal{L}_1^2(k) & =
    \begin{pmatrix}
      \bbraket{k, + | \mathcal{L}_1^2 | k, +} & \bbraket{k, + | \mathcal{L}_1^2
        | k, -} \\ \bbraket{k, - | \mathcal{L}_1^2 | k, +} & \bbraket{k, - | \mathcal{L}_1^2 | k, -}
    \end{pmatrix}
    \\ & = 2 J^2
    \begin{pmatrix}
      1 & - e^{ik} \\
      -e^{-ik} & 1
    \end{pmatrix}
    , \\
    \mathcal{L}_3^2(k) & = 2 J^2
    \begin{pmatrix}
      1 & - e^{-ik} \\
      -e^{ik} & 1
    \end{pmatrix}
    , \\
    \mathcal{L}_2^2(k) & = \mathcal{L}_4^2(k) = 2 J^2 \left( \id - \sigma_x
    \right),
  \end{split}
\end{equation}
where $\sigma_x$ is the Pauli matrix. It is straightforward to obtain the
Floquet superoperator in Eq.~\eqref{eq:F-alternative-noisy-Floquet-topo-chain}
(keeping in mind that $\mathcal{U}_{\rm F} = \id$) and to diagonalize it. Writing the
eigenvalues as $e^{-i \lambda(k)}$, we obtain two bands which we denote by
$\lambda_{\rho}(k)$ and $\lambda_{\sigma}(k)$, respectively,
\begin{multline}
  \label{eq:eq:lambda-rho-sigma}
  e^{-i \lambda_{\rho, \sigma}(k)} = \frac{1}{16} e^{-8 \kappa ^2} \left\{ 10
    e^{4 \kappa ^2}+3 e^{8 \kappa ^2} + 3 \vphantom{\left[ \left(e^{4 \kappa
            ^2}-1\right)^2 \right]^{1/2}} \right. \\ - \left( 8 e^{4 \kappa ^2}
    - 4 e^{8 \kappa ^2} - 4 \right) \cos(k) - \left( 2 e^{4 \kappa ^2} - e^{8
      \kappa ^2} - 1 \right) \cos(2 k) \\ \pm 2 \sqrt{2} \left(e^{4 \kappa ^2} -
    1 \right) \cos(k/2)^2 \left[ 26 e^{4 \kappa ^2}+3 e^{8 \kappa ^2} + 3
    \vphantom{\left(e^{4 \kappa ^2}-1\right)^2} \right. \\ \left. \left. + 4
      \left(e^{4 \kappa ^2}-1\right)^2 \cos (k)+\left(e^{4 \kappa ^2}-1\right)^2
      \cos (2 k) \right]^{1/2} \right\}.
\end{multline}
Both $\lambda_{\rho}(k)$ and $\lambda_{\sigma}(k)$ are purely imaginary, and the
imaginary parts encode the decay rates of the respective eigenoperators. For
$k \to 0$, $\lambda_{\rho}(k)$ goes to zero quadratically. Thus, the
corresponding eigenoperators are long-lived. In contrast,
$\Im(\lambda_{\sigma}(k)) < 0$ for all values of $k$, so that the eigenoperators
corresponding to this band decay quickly. These properties can be seen most
clearly in the limit of weak timing noise, in which the expressions for
$\lambda_{\lambda}(k)$ and $\lambda_{\sigma}(k)$ simplify considerably:
\begin{equation}
  \label{eq:lambda-rho-sigma-weak-noise}
  \begin{split}
    \lambda_{\rho}(k) & = - i 2 \kappa^2 \left( 1 - \cos(k) \right) + O(\kappa^4), \\
    \lambda_{\sigma}(k) & = - i 2 \kappa^2 \left( 3 + \cos(k) \right) + O(\kappa^4).
  \end{split}
\end{equation}
The eigenoperators corresponding to the two bands take the form
\begin{equation}
  \label{eq:rho-sigma-superkets}
  \begin{split}
    \kket{\rho(k)} & = \frac{1}{\sqrt{2}} \left( \kket{k, +} + e^{i \phi_{\rho}(k)} \kket{k, -}
    \right), \\ \kket{\sigma(k)} & = \frac{1}{\sqrt{2}} \left( \kket{k, +} -
      e^{i \phi_{\sigma}(k)} \kket{k, -} \right),
  \end{split}
\end{equation}
i.e., both have equal weight on the states $\kket{k, +}$ and $\kket{k, -}$, with
relative phases between these states that depend on the momentum $k$. At low
momenta, $k \to 0$, or weak timing noise, $\kappa \to 0$, both phases vanish,
$\phi_{\rho}(k), \phi_{\sigma}(k) \to 0$. Note that we obtain the same structure for the
left eigenvectors of $\mathcal{F}$.

These results allow us to calculate the survival probability of localized states
analytically. For simplicity, we consider an initial state that is an incoherent
superposition of a particle localized on the $+$ and $-$ sublattice sites in the
middle of the chain at $j = 0$, i.e.,
$\kket{\rho_0} = \frac{1}{2} \left( \kket{0, +} + \kket{0, -} \right)$, where
the normalization is chosen such that $\tr(\rho_0) = 1$. The overlaps of this
density matrix with the left eigenoperators corresponding to the bands
$\lambda_{\rho}(k)$ and $\lambda_{\sigma}(k)$ are
$\bbraket{\rho(k) | \rho_0} = 1/\sqrt{2}$ and
$\bbraket{\sigma(k) | \rho_0} = 0$.  Moreover, $\kket{\rho_0}$ has vanishing
overlap with any ``off-diagonal'' operators such as $\ket{j, s} \bra{j', s'}$.
Hence, the survival probability Eq.~\eqref{eq:survival_superoperator} is given by,
\begin{equation}
  \label{eq:survival-k-integral}
  P_{\mathrm{s}} = \frac{1}{2} \int_{-\pi}^{\pi} \frac{d k}{2 \pi} e^{-i n \lambda_{\rho}(k)}. 
\end{equation}
For $n \to \infty$, the integral over momenta can be evaluated in a
stationary-phase approximation. To this end, we expand $\lambda_{\rho}(k)$ in
the vicinity of its minimum at $k = 0$,
\begin{equation}
  \lambda_{\rho}(k) = -\frac{k^2}{2} \tanh(2 \kappa^2) + O(k^4).
\end{equation}
Then, extending the integration over momenta to the full real line, we find
\begin{equation}
  s \sim \frac{1}{\sqrt{8 \pi n \tanh(2 \kappa^2)}},
\end{equation}
i.e., diffusive decay for any value of $\kappa$. At weak noise, we can expand
the hyperbolic tangent in $\kappa$ and recover the result of
Ref.~\cite{Rieder2017}. Actually, for $\kappa \to 0$, a closed expression for
the survival probability can be obtained over the full range of $n$ by using
Eqs.~\eqref{eq:lambda-rho-sigma-weak-noise}. Then, the integral in
Eq.~\eqref{eq:survival-k-integral} yields a modified Bessel function,
\begin{equation}
  \label{eq:survival-bessel}
  P_{\mathrm{s}} = \frac{1}{2} \int_{-\pi}^{\pi} \frac{d k}{2 \pi} e^{- 2 n
    \kappa^2 \left( 1 - \cos(k) \right)} = \frac{1}{2} e^{- 2 n \kappa^2} I_0(2
  n \kappa^2). 
\end{equation}
As shown in Fig.~\ref{fig:survivals}, this result is in excellent agreement with
the numerically determined survival probability. We finally note that using the
asymptotic expansion of the Bessel function, $I_0(x) \sim e^x/\sqrt{2 \pi x}$
for $x \to \infty$, leads again to diffusive decay,
$P_{\mathrm{s}} \sim 1/(4 \kappa \sqrt{\pi n})$ for $n \to \infty$.

Our analytical treatment of the noisy Floquet topological chain at resonant
driving shows that the diffusive decay of the end state can be traced back to
two key properties of the eigenoperators and eigenvalues of the Floquet
superoperator: (i) The end state has spectral weight on a continuum (in the
thermodynamic limit) of eigenoperators which form the band $\lambda_{\rho}(k)$,
and (ii) the corresponding eigenvalues accumulate at $\lambda = 0$. More
specifically, the quadratic vanishing of $\lambda_{\rho}(k)$ for $k \to 0$
implies that the ``density of states'' $\sim 1/(d \lambda_{\rho}(k)/dk)$ has a
square-root divergence at $\lambda = 0$.  These properties, together with the
expression for the survival probability Eq.~\eqref{eq:survival_superoperator},
provide an intuitive way of understanding the form of the the end mode decay
shown in Fig.~\ref{fig:survivals}: The survival probability is obtained by
integrating over a continuum of real exponential functions, leading to a
diffusive decay.

Figures~\ref{fig:Fevals_circle} and \ref{fig:Fevals_imag} show the numerical check of this intuition. 
In the right panel of Fig.~\ref{fig:Fevals_circle}, which pertains
to resonant driving, we observe that indeed a large number of eigenoperators of
${\cal F}$ have a nonzero overlap with the edge superket. In fact, as
illustrated in the bottom panel of Fig.~\ref{fig:Fevals_imag}, the spectral weight
of the edge superket is almost evenly distributed over the set of states forming
the band $\lambda_{\rho}(k)$ identified in our analytical approach. This
confirms property (i). A first indication of property (ii) can also be seen in
the main panel of Fig.~\ref{fig:Fevals_imag} -- evidently, eigenvalues of
$\mathcal{F}$ that have non-zero overlap with $\kket{e}$ cluster at
$\Im(\lambda) = 0$. To demonstrate this feature unambiguously, we plot in the
inset of Fig.~\ref{fig:Fevals_imag} the imaginary parts of these eigenvalues in
decreasing order and find good agreement with the analytical prediction in
Eq.~\eqref{eq:lambda-rho-sigma-weak-noise}. This in turn implies that the
density of states has a square-root singularity at $k \to 0$, which is rounded
in the numerics due to the finite system size.

\subsubsection{Diffusive decay for a disorder-localized bulk}
\label{sec:diff-decay-disord}

As explained above and illustrated in Fig.~\ref{fig:survivals}, the decay of the
end state survival probability is diffusive if the bulk states are
localized. Localization can be due to resonant driving conditions -- we
considered this fine-tuned situation, which is amenable to a fully analytical
treatment, in the previous section. In the more generic case of driving
parameters that do not match the resonance condition, localization of the bulk
states can be induced by adding disorder to the system, i.e., by modifying the
driving protocol Eq.~\eqref{eq:Hamiltonian-Floquet-topo-chain} as
\begin{equation}
  \label{eq:Hamiltonian-Floquet-topo-chain-dis}
  H(t) = H_i + H_{\mathrm{dis}} \quad \text{for} \quad \left( i-1 \right) T/4 \leq t < iT/4,
\end{equation}
where $H_i$ and $H_{\mathrm{dis}}$ are given in
Eqs.~\eqref{eq:Hamiltonian-step-Floquet-topo-chain} and~\eqref{eq:Hdis},
respectively. The quenched disorder term $H_{\mathrm{dis}}$ does not commute
with the clean Hamiltonians $H_i$. Therefore, we form superoperators
${\cal H}_i$ from the full disordered Hamiltonian during each driving step,
$H_i + H_{\rm dis}$. This allows us to perform the noise average separately for
each quenched disorder realization, and leads again to a Floquet superoperator
of the form of Eq.~\eqref{eq:F-noisy-Floquet-topo-chain}, which factorizes into
contributions corresponding to the individual driving steps.

While the inclusion of disorder complicates analytics significantly, we can
still gain valuable insight from the Floquet superoperator formalism if we
determine the spectrum of $\mathcal{F}$ numerically. In particular, above we
identified a square-root singularity of the density of states at $\lambda = 0$
as the root cause for the diffusive decay. Here, we provide evidence that this
phenomenon persists in a disordered system.

\begin{figure}
 \includegraphics[width=\columnwidth]{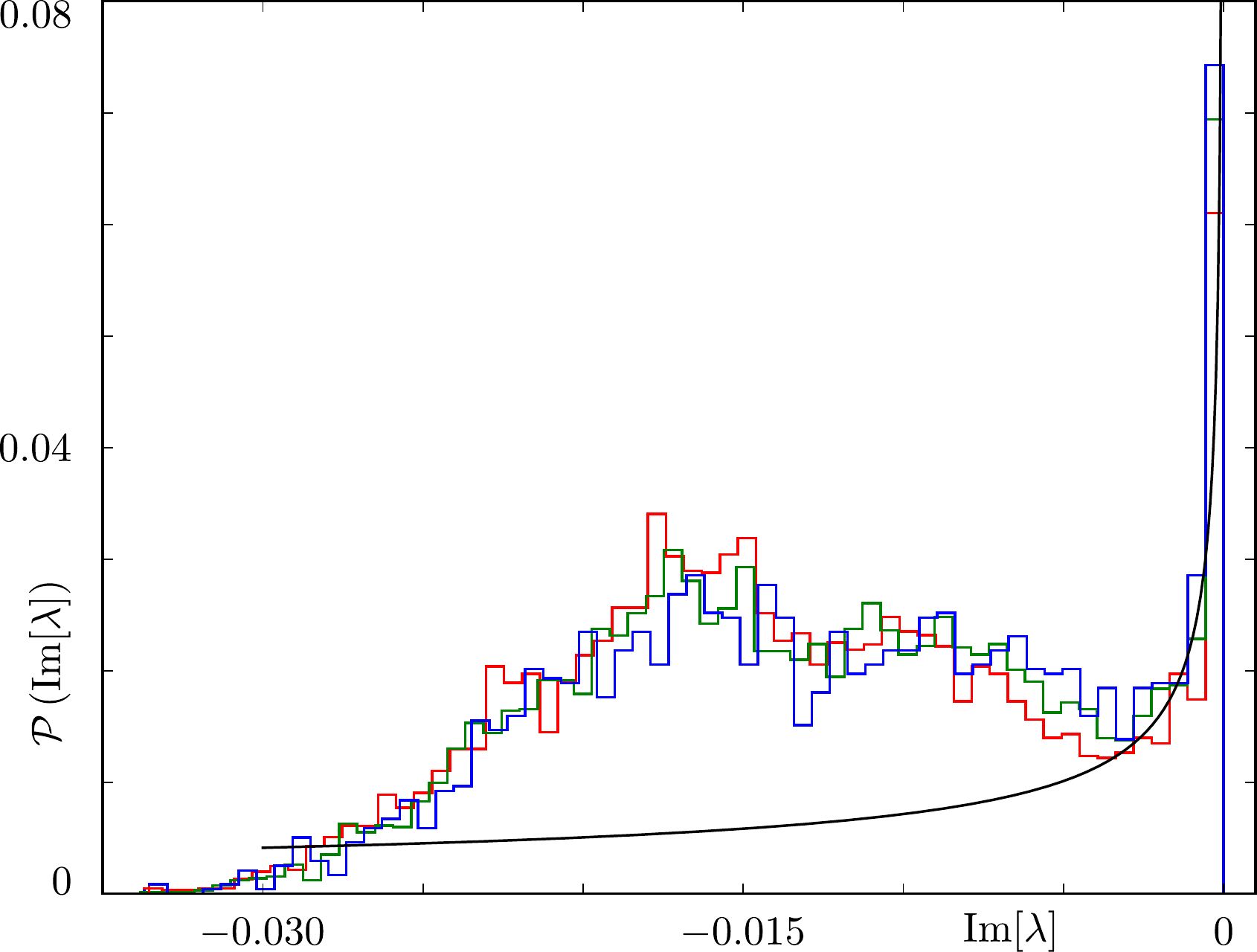}
 \caption{Probability distribution of ${\rm Im}[\lambda]$, ${\cal P}({\rm Im}[\lambda])$,  corresponding to all
   eigenoperators of ${\cal F}$ for which the squared overlap with the edge superket is
   greater than $10^{-4}$. We use an on-site disorder strength $VT/4=0.4$ and a
   noise strength $\tau/T=1/80$. The three histograms represent different system
   sizes: 40 rungs (red), 50 rungs (green), and 60 rungs (blue). They are
   obtained from 150 simultaneous disorder-noise realizations in the case of 40
   and 50 rung systems, and by using 50 realizations for the largest system
   size. The probability distribution is peaked towards $\lambda\to0$, with a
   peak hight that increases with system size. The black curve shows a square
   root singularity, which by Eq.~\eqref{eq:survival_superoperator} leads to a
   diffusive decay of the end mode survival
   probability. \label{fig:Fevals_distro}}
\end{figure}

Figure.~\ref{fig:Fevals_distro} shows the probability distribution of
${\rm Im}[\lambda]$ for all states with an overlap
$|\bbraket{\alpha|e}|^2>10^{-4}$. When the diffusive decay results from adding disorder to the system, the
eigenvalues of ${\cal F}$ take random values for each disorder realization.
However, we observe that the distribution is peaked
towards $\lambda\to0$, signifying that there is a large number of slowly
decaying states. The peak height grows with increasing system size for the sizes
we can access numerically, and its decay away from $\lambda=0$ is comparable to
$1/\sqrt{-{\rm Im}[\lambda]}$ (black line). Given the form of the survival
probability Eq.~\eqref{eq:survival_superoperator}, a square root singularity
leads to a diffusive decay, consistent with Fig.~\ref{fig:survivals}.

\section{Conclusions and Outlook}
\label{sec:conclusions-outlook}

In this paper, we discussed a general formalism to treat randomly driven quantum
systems with periodically recurring statistics. Our focus was on a description
of the averaged system dynamics. We showed that the average over random
fluctuations of the drive leads to an evolution equation of the system's density
matrix in terms of a Floquet superoperator~\eqref{eq:caligraphic-F}. The
eigenoperators and eigenvalues of the Floquet superoperator are generalizations
of Floquet states and quasienergies, which are familiar from the theory of
periodically driven systems without noise, to the randomly driven setting. In
particular, a spectral representation of the Floquet superoperator in terms of
its eigenoperators and their corresponding eigenvalues enables an efficient
description of the system dynamics in analogy to the spectral representation of
the usual Floquet operator in terms of Floquet states and quasienergies.

We developed this formalism in the context of noisy Floquet systems and showed
that it can also be applied in systems with fully random driving. To illustrate
the formalism, we revisited a model of a Floquet topological chain which we
previously studied in Ref.~\cite{Rieder2017}, and we re-derived and corroborated
our earlier results. Our analysis of the spectrum of the Floquet superoperator
revealed that the diffusive decay of the end state of a Floquet topological
chain with a localized bulk can be traced back to an accumulation of long-lived
modes, i.e., eigenvalues with vanishing imaginary part.

In this work, we focused on piecewise constant driving protocols, for which
timing noise is a naturally present source of unwanted random
fluctuations. Other types of noise have to be taken into account for continuous
driving as is realized, e.g., in laser-induced Floquet topological insulators in
condensed matter experiments~\cite{Wang2013a}. In particular, for the simple yet
ubiquitous case of harmonic driving described by
$H(t) = H_0 + H_1 A \sin(\omega t + \phi)$ with time-independent Hamiltonians
$H_0$ and $H_1$, noise might affect the amplitude $A$, frequency $\omega$, and
phase $\phi$ of the drive. Which type of noise dominates depends both on the
concrete experimental realization and the physical phenomenon under study. From
the perspective of theory, an immediate question is whether such a situation can
also be described by a Floquet superoperator formalism. We expect that this is
the case if the noise is Markovian, i.e., correlations of the noise decay much
faster than all other relevant time scales set by the Hamiltonians $H_0$ and
$H_1$, as well as the period of the drive. This expectation is grounded on the
observation that our derivation of a formal expression for the Floquet
superoperator relied solely on the assumption of statistical independence of
fluctuations in different driving cycles, which translates to assuming Markovian
correlations for temporally continuous noise processes. The precise form of the
Floquet superoperator for amplitude, frequency, and phase noise, and the
physical consequences of such types of noise for, e.g., the decay of edge
states, are interesting questions for further studies.

Above, we applied the Floquet superoperator formalism to a one-dimensional noisy
Floquet topological chain, and it will be interesting to extend this analysis to
higher-dimensional Floquet topological phases. As in the one-dimensional case,
noise will lead to leakage from topological surface modes into the
bulk. However, in contrast to the Floquet topological chain considered here,
surface modes in higher-dimensional systems are propagating, and it is an open
question how the propagation along the system's surface will be affected by
different types of noise.

\begin{acknowledgments}
  We warmly thank C.~Roberto for insightful discussions, as well as Ulrike
  Nitzsche for technical assistance. LMS acknowledges support by the ERC through
  the synergy grant UQUAM and MTR acknowledges support by the Alexander von
  Humboldt Foundation.
\end{acknowledgments}

\appendix
\section{Implementation of superoperators}
\label{app:super}

To study the properties of the noise-averaged Floquet superoperator
$\mathcal{F}$ it is convenient to obtain a matrix
representation~\cite{Zwolak2004}. For this purpose we regard density matrices as
vectors, ``superkets,'' in the space of operators. Starting from the
lattice-site basis of states $\ket{j, s}$ of the original Hilbert space in the
example of Sec.~\ref{sec:appl-noisy-floq}, a basis of operator space is given by
\begin{equation}
  \kket{j, s; j', s'} = \ket{j, s} \bra{j', s'}.
\end{equation}
A scalar product in the space of operators is defined by
$\bbraket{A | B} = \tr(A^{\dagger} B)$, which
implies
\begin{multline}  
  \bbraket{j_1, s_1; j_2, s_2 | j_3, s_3; j_4, s_4} \\
  \begin{aligned}
    & = \tr( \ket{j_2, s_2} \braket{j_1, s_1 | j_3, s_3} \bra{j_4, s_4} ) \\ & =
    \delta_{j_1 j_3} \delta_{s_1 s_3} \delta_{j_2 j_4} \delta_{s_2 s_4}.
  \end{aligned}
\end{multline}
In the basis of operators $\kket{j, s; j', s'}$, the matrix elements of the
superoperator $\mathcal{H}_i$, which is defined by its action on an operator
$A$, $\mathcal{H}_i A = [H_i, A]$, are thus given by
\begin{multline}
  \label{eq:H_i_matrix_elements}
  \bbraket{j_1, s_1; j_2, s_2 | \mathcal{H}_i | j_3, s_3; j_4, s_4} \\
  \begin{aligned}
    & = \tr( \ket{j_2, s_2} \bra{j_1, s_1} [H_i, \ket{j_3, s_3} \bra{j_4, s_4}]
    ) \\ & = \bra{j_1, s_1} [H_i, \ket{j_3, s_3} \bra{j_4, s_4}] \ket{j_2, s_2}
    \\ & = \braket{j_1, s_1 | H_i | j_3, s_3} \delta_{j_4 j_2} \delta_{s_4 s_2}
    \\ & \peq - \delta_{j_1 j_3} \delta_{s_1 s_3} \braket{j_4, s_4 | H_i | j_2, s_2}.
  \end{aligned}
\end{multline}
With the matrix representation of the superoperators $\mathcal{H}_i$ for
$i = 1, 2, 3, 4$, it is straightforward to assemble the Floquet superoperator
$\mathcal{F}$ [Eq.~\eqref{eq:F-noisy-Floquet-topo-chain}] for the noisy Floquet
topological chain discussed in Sec.~\ref{sec:appl-noisy-floq}. All numerical
results presented in that section are based on the exact diagonalization of the
Floquet superoperator.


%

\end{document}